\documentclass[a4paper,11pt]{article}
\pdfoutput=1
\usepackage{jheppub}
\usepackage{amssymb,amsmath,mathrsfs,enumerate}
\usepackage{graphicx,rotate,multicol}
\usepackage{float}

\usepackage{subfig}

\usepackage{slashed}
\usepackage{mathtools}
\usepackage{multirow}

\allowdisplaybreaks

\title{\boldmath Bounds on Exotic Couplings from a New $\nu$-Background}

\author{Indra Kumar Banerjee,}
\author{Ujjal Kumar Dey,}
\author{Anna John}
\affiliation{Department of Physical Sciences, Indian Institute of Science Education and Research Berhampur,\\ Ganjam, Odisha 760003, India}

\emailAdd{indrab@iiserbpr.ac.in}
\emailAdd{ujjal@iiserbpr.ac.in}
\emailAdd{anna23@iiserbpr.ac.in}

\abstract{We propose a hitherto unexplored neutrino background emerging from the mechanism of quenched superradiance of rotating primordial black holes. The quenching of the phenomenon happens through fermionic production, in our case neutrino production, from the boson cloud formed due to superradiance. The couplings involved in these interactions are bounded from above through several studies. In this work we put lower bounds on such scalar and vector couplings.
}

\begin{document}
\maketitle
\flushbottom

\section{Introduction}
\label{sec:intro}
Primordial black holes (PBH), formed in the early universe, due to the gravitational collapse of extremely dense regions, can be a laboratory for many exotic events which could give a detectable flux of standard model (SM) particles, beyond the standard model (BSM) particles, and also gravitational waves (GW). Unlike astrophysical black holes (ABH), PBHs can have very low masses and they can be viable dark matter candidate under certain parameter choices \cite{Carr:2020gox}. Similar to ABHs, PBHs are also characterized by their mass and spin.
Superradiance of spinning black holes \cite{Starobinskii:1973vzb, Brito:2015oca,Press:1972zz,Detweiler:1980uk,Zouros:1979iw,1971ZhPmR..14..270Z} is a phenomenon which has been vastly studied and various efforts have been made to detect signals arising from them \cite{Arvanitaki:2010sy,Fernandez:2019qbj,Branco:2023frw,Calza:2023rjt,East:2017ovw,Baryakhtar:2017ngi,Ghosh:2021uqw,Banerjee:2023vst,Chung:2021roh,Bernal:2022oha,Jia:2025vqn,Ferraz:2020zgi,Rosa:2017ury}. In superradiance, bosonic particles are spontaneously created and the bosonic field undergo amplification by angular momentum extraction from the rotating black holes. They form a bosonic cloud around the black hole \cite{Bekenstein:1973mi,Bekenstein:1998nt}. Such a system is often called a \textit{``black hole bomb"} due to the unstable amplification of the cloud. The growth advances till the spin of the black hole drops below a threshold value and superradiance stops. This is the usual case of superradiance which is widely studied. Recently, a particular case of this phenomenon was introduced in \cite{Chen:2023vkq} where it has been suggested that if there is an adequate coupling between the bosons and other fermions this instability can be dealt efficiently. As the cloud extracts energy from the black hole an equal amount of energy can be extracted from the cloud through interaction with the fermions, creating detectable fermion fluxes which in turn \textit{quenches} the growth of the cloud. This mechanism allows superradiance to go on for a much longer time compared to the vanilla scenario. 
In this article we focus on quenched superradiance of spinning primordial black holes considering that the BSM bosons that facilitate the mechanism, possess some coupling with the neutrinos. There are various BSM scenarios where a coupling between BSM bosons and neutrinos (active and sterile) has been considered and till now some upper bounds on the coupling have been obtained \cite{Farzan:2002wx, Forastieri:2019cuf, Kolb:1987qy, Kachelriess:2000qc, Fiorillo:2022cdq, Abdullah:2018ykz, Cadeddu:2020nbr, Venzor:2020ova,Cline:1993ht,Pilaftsis:1993af,Chang:1993yp,Ivanez-Ballesteros:2024nws,Bilmis:2015lja,Lindner:2018kjo,Sevda:2017wwn,Gninenko:2020xys,Laha:2013xua,Bonilla:2023dtf}. The couplings in these cases are bounded from above taking into consideration cosmological, astrophysical and experimental observations. The coupling between neutrinos and the bosons can modify $N_{\text{eff}}$ which affects the expansion rate of the universe and consequently the timeframe of big bang nucleosynthesis (BBN) \cite{Wagoner:1966pv,Dolgov:2002wy}. This in turn give bounds on the coupling by comparing the predicted and observed light element abundances. Other studies deal with neutrino and boson (majoron) coupling in the core of supernova and these bosons contribute to an additional energy loss. There are also cases where due to these couplings neutrinos of a higher energy range are emitted. Due to the non-observation of these in the SN1987A data the couplings are highly constrained \cite{Kamiokande-II:1987idp, Hirata:1988ad, Koshiba:1992yb, Oyama:2021oqp}. Some of these works use the coherent elastic neutrino-nucleus scattering (CE$\nu$NS) process measurement by the COHERENT Collaboration to obtain these upper bounds \cite{COHERENT:2017ipa}. In our study, for the first time, we constrain these couplings from below, that is, we provide lower bounds for these couplings with respect to the mass of the boson. In order to do so, we make two simple assumptions, i.e., PBHs with (i) high initial spin \cite{Saito:2023fpt,Kuhnel:2019zbc}, and (ii) the `right' mass \cite{Carr:2020gox} make up a significant part of the dark matter of the universe. The `right' PBH mass, corresponding to the mass of the boson, allows for an efficient superradiance process if the PBHs have sufficient initial spin. Since these PBHs originate in the very early phases of the Universe, they can go through quenched superradiance to create neutrinos. Furthermore, since we consider the PBHs to make up a significant amount of dark matter, we consider the creation of these neutrinos throughout the Universe. This, in principle, can give rise to a novel diffuse neutrino background from the quenched superradiance (DQS$\nu$B) process which leads us to the aforementioned bounds.
This article is structured as follows. In Sec~\ref{sec:qsr} we provide a brief overview of the quenched superradiance mechanism followed by Sec.~\ref{sec:dqsnb} where we discuss the diffuse neutrino background originated from the superradiance mechanism. In Sec.~\ref{sec:res} we show our results, and finally in Sec.~\ref{sec:concl} we summarize and conclude.

\section{Quenched Superradiance Mechanism}
\label{sec:qsr}
All black holes possess a characteristic one way membrane, a point of no return, namely the \textit{event horizon}. Moreover, in rotating black holes there is a region surrounding the event horizon, called the ergoregion, where no object can remain stationary with respect to an outside observer due to the Lense-Thirring effect resulting in corotation of objects with the black hole. Unlike the event horizon, in the ergoregion, particles can have an escape velocity high enough to evade the gravitational pull of the black hole. Thus it is possible for objects to enter this region, extract energy from the rotating black hole, and exit with a higher energy. This idea led to the conception of the well-known Penrose process where an object with a certain positive definite energy enters the ergoregion and from it a higher energy and a negative energy state are created. The negative energy state falls into the black hole while the state with energy higher than the initial state comes out. Superradiance is an analogous process for fields. Bosonic particles can be spontaneously generated due to the amplification of the corresponding field through the extraction of rotational energy from the compact object. These bosonic particles then form a cloud-like structure around it. This creates a ``\textit{gravitational atom}" where the bosonic cloud is gravitationally bound by the black hole which acts as the nucleus. The gravitational atom can be characterized by the gravitational fine structure constant  $\alpha_{g}=GM_{\text{BH}}m_{\psi}/(\hbar c)$, where $G$ is Newton's gravitational constant, $M_\text{{BH}}$ the mass of the black hole and $m_{\psi}$ the mass of the boson \cite{Brito2015Superradiance}.
The gravitational fine structure constant, $\alpha_{g}$ signifies the efficiency of the energy extraction and hence the superradiance process has significant dependence on it.
The ground states of the scalar and vector bosonic fields around a rotating compact object can be expressed in the Boyer-Lindquist coordinates ($t,r,\theta,\phi$) as \cite{Brito2015Superradiance},
\begin{subequations}
\begin{align}
	\label{scalarfield}
    \Phi(t)&=\Psi_{0}(t)e^{1-\alpha_{g}^{2}r/2r_{g}}\frac{\alpha_{g}^{2}r}{2r_{g}}(\alpha_{g}\sin\theta\cos(m_{\Phi}t-\phi),\\
\label{vectorfield}
    A^{\prime \, \mu}(t)&=\Psi_{0}(t)e^{-\alpha_{g}^{2}r/r_{g}}\left(\alpha_{g}\sin\theta\sin(m_{A^{\prime}} t - \phi), \cos(m_{A^{\prime}} t), \sin(m_{A^{\prime}} t), 0\right),
\end{align}
\end{subequations}
respectively, where $m_{\Phi}$ and $m_{A^{\prime}}$ are the masses of the scalar and the vector bosons respectively, $r_{g}=GM_{\text{BH}}$ is the gravitational radius, and $\Psi_{0}$ is the peak field value.
The mass of the cloud and the peak field value are connected by the following expressions \cite{Chen:2023vkq},
\begin{equation}	
M_{\text{c}} =
\begin{cases}
  \frac{186 \Psi_{0}^{2}}{\alpha_{g}^{3}m_{\phi}},& \text{scalar}, \\  
    \frac{\pi \Psi_{0}^{2}}{\alpha_{g}^{3}m_{A^{\prime}}}, & \text{vector}.
\end{cases}
\label{fieldmass}
\end{equation}
It is to be noted that as the bosonic cloud grows, it extracts more energy from the BH and hence the growth becomes faster and the rate of energy extraction also increases. This goes on till the remaining rotational energy can not sustain the superradiance process any more.
This condition can be mathematically expressed as,
\begin{equation}
\alpha_{g}<\frac{\alpha_{J}}{2\left(1+\sqrt{1-\alpha_{J}^{2}}\right)},
\label{spin}
\end{equation}
where $\alpha_{J}$ is the spin of the black hole.
Roughly it can be estimated that when the mass of the cloud goes beyond $10\%$ of $M_{\text{BH}}$ the spin of the black hole decreases and the rotational energy of the black hole will no longer be sufficient for superradiance to occur \cite{Chen:2023vkq}. 
In such a case the maximum field value when the cloud mass reaches the threshold is 
\begin{equation}
\Psi_{10\%} =
\begin{cases}
    1.1\times10^{25}\left(\frac{\alpha_{g}}{0.2}\right)^{2}  \text{eV}, & \text{scalar}, \\
   8.7\times10^{25}\left(\frac{\alpha_{g}}{0.2}\right)^{2} \text {eV},& \text{vector}.
\end{cases}
\label{psimax}
\end{equation}
It is worth mentioning here that since the growth of the bosonic cloud positively influences the energy extraction which in turn positively effects the growth, superradiance can diminish the spin of a rotating black hole very fast.
\subsection{Quenching of Superradiance: Production of Neutrinos}
\label{subsec:nuprod}
So far the discussion has been around a run-of-the-mill scenario of superradiance. As mentioned in Eq.~\eqref{spin}, there is a threshold value of the spin of the black hole below which superradiance dies down. This is due to the continuous extraction of energy from the black hole by the boson cloud. However, it is intuitive to argue that, if there is a sufficient amount of energy loss from this cloud, superradiance mechanism could be sustained for longer. As explored in \cite{Banerjee:2024knt,Chen:2023vkq} the prolongation of the superradiance phenomenon is facilitated by the interaction of the bosons with fermions around them. In such a case the energy dissipated from the cloud due to the interaction with the fermions will be an offset against the energy gained by the cloud. This establishes a superradiance balance as the cloud mass stays the same for a much longer time as compared to the vanilla case keeping the energy extraction in check. This mechanism is termed ``quenched superradiance" \cite{Banerjee:2024knt}. The fermion production depends on the boson that participates in the superradiance process, i.e., fermions are produced through (i) paramateric excitation of oscillating scalar fields \cite{Greene:1998nh,Greene:2000ew}, and (ii) Schwinger pair production from vector fields \cite{Schwinger:1951nm}.
In our study we take the fermions to be neutrinos and investigate the outcome for this scenario.
In the case of the scalar field the generic interaction between the fermion and the boson is described as $g_{s}\phi \overline{f}f$. Since we consider the fermions to be neutrinos, there are cosmological constraints on the coupling $g_s$ from BBN \cite{Venzor:2020ova,Huang:2017egl}, CMB \cite{Forastieri:2019cuf,Escudero:2019gvw,Sandner:2023ptm,Li:2023puz} etc. whereas constraints from other sources come from neutrinoless double beta decay \cite{Blum:2018ljv,KamLAND-Zen:2012uen,Brune:2018sab}. These constraints put an upper bound roughly around $10^{-7}$.  It is important to note that for the neutrino production from scalars to occur the field values must be high enough and should obey the condition $g_{s}\Psi_{0}\gg m_{\nu}$. However, a coupling between the scalar field and the active neutrinos would not provide an efficient channel to maintain the balance since the fluctuating scalar field can cause the neutrinos to acquire a very high effective mass which would then lead to their decay into pions after production. Thus for the scalar case, we take the fermions to be sterile neutrinos of mass $\mathcal{O}(1~\text{eV})$ which would oscillate into active neutrinos during propagation. The rate of neutrino production through this parametric excitation is given as,
\begin{equation}
    \Gamma_{s}=\frac{g_{s}^{2}\phi_{0}^{2}m_{\phi}^{2}}{48 \pi^{3}}\sqrt{\frac{m_{\phi}}{g_{s}\phi_{0}}}~ .
    \label{paraexcitrate}
\end{equation}
In the case of vector fields, the interaction between the boson field and the fermion takes the form $g_{v}A^{\prime \mu}\overline{f}\gamma_{\mu}f$. The Schwinger pair production of neutrinos from the vector bosons requires that the condition $g_{v}E_{A^{\prime}}\gg m_{f}^{2}$ is satisfied where $E_{A^{\prime}}\approx m_{A^{\prime}} |\vec{A^{\prime}}|$. The production rate is given by,
\begin{equation}
    \Gamma_{v}=\frac{g_{v}^{2}E_{A^{\prime}}^2}{48 \pi}.
    \label{schwprodrate}
\end{equation}
The neutrinos produced from the scalar and vector fields are immediately accelerated to higher energies which lead to a detectable flux of these high energy neutrinos. The energy and the differential flux of these neutrinos from a single PBH source are respectively given as \cite{Chen:2023vkq},
\begin{equation}
E_{f} \approx
\begin{cases}
    0.27 \, g_s \Psi_0, & \text{scalar,} \\
    0.35 \, g_v \Psi_0, & \text{vector,}
\end{cases}
\label{energy}
\end{equation}
and
\begin{small}
\begin{align}
\frac{d\phi_{f}}{dE} \approx
\begin{cases}
    1.2 \times 10^{-17} \, \text{cm}^{-2} \, \text{s}^{-1} \, \text{eV}^{-1} \left(\frac{\Psi_0/\text{GeV}}{4.8 \times 10^{7}}\right)^{1/2} \left(\frac{N_{f}}{3}\right) \left(\frac{10^{-12}}{m_{\phi}/\text{eV}}\right) \left(\frac{\alpha}{0.3}\right)^3 \left(\frac{g_s}{10^{-8}}\right)^{1/2} \left(\frac{5}{d/\text{kpc}}\right)^2, ~ \text{scalar,} \\
    1.3 \times 10^{-6} \, \text{cm}^{-2} \, \text{s}^{-1} \, \text{eV}^{-1} \left(\frac{\Psi_0/\text{GeV}}{5.7 \times 10^{14}}\right) \left(\frac{N_{f}}{1}\right) \left(\frac{10^{-12}}{m_{A^{\prime}}/\text{eV}}\right) \left(\frac{\alpha}{0.3}\right)^3 \left(\frac{g_v}{10^{-12}}\right) \left(\frac{5}{d/\text{kpc}}\right)^2, ~~ \text{vector,}
\end{cases}
\label{difflux}
\end{align}
\end{small}
where $N_{f}$ is the number of fermion (neutrino) mass or flavor eigenstates taken into consideration and $d$ is the distance of the BH from the observer in kpc. It is worth mentioning here that the above expression for the neutrino flux due to the scalar case is for the sterile neutrino flux. In order to obtain the active neutrino flux for this case, we consider the active-sterile mixing $\mathcal{O}(10^{-3})$ and for the subsequent parts of the article we take this mixing into account in all the calculations and figures.
As mentioned before, neutrino production may lead to superradiance balance. During this phase, i.e., where the energy extracted by the bosonic cloud is same as the energy extracted from the cloud, the superradiant growth rate and neutrino production rate shall be equated and we obtain a critical value of the maximum field value~\cite{Chen:2023vkq}, 
\begin{align}
\Psi_{0}^{\text{c}} \approx
\begin{cases}
    4.8 \times 10^{16}\left(\frac{3}{N_{f}}\right)^{2}\left(\frac{m_{\phi} \text{/eV}}{10^{-12}}\right)\left(\frac{\alpha_{g}}{0.3}\right)^{16}\left(\frac{10^{-7}}{g_{s}}\right)^{5}\left(\frac{\alpha_{J}}{0.9}\right) \text{eV}, & \text{scalar},\\
    5.7 \times 10^{14}\left(\frac{1}{N_{f}}\right)\left(\frac{m_{A^{\prime}} \text{/eV}}{10^{-12}}\right)\left(\frac{\alpha_{g}}{0.3}\right)^{6}\left(\frac{10^{-12}}{g_{v}}\right)^{3}\left(\frac{\alpha_{J}}{0.9}\right) \text{eV},&\text{vector}.
\end{cases}
\label{psicrit}
\end{align}
This critical value of the field depends on the gravitational fine structure constant, the coupling between the fermion and the boson and also on the mass of the boson. It is to be noted here that although we mentioned the balanced phase, there are a few other relevant phases during the lifetime of the quenched superradiance which we discuss below.

\subsection{Timeline of Quenched Suprradiance}
\label{subsec:timeqsr}
The phases depending on the flow of energy in quenched superradiance is divided into three phases, the superradiant growth phase, the balance phase, and finally the depletion phase. We discuss them briefly in the following. During these phases there is an evolution of the mass of the cloud ($M_{\text{c}}$) with respect to time, which is given by \cite{Banerjee:2024knt},
\begin{equation}
\frac{dM_{\text{c}}}{dt} = \Gamma_{\text{SR}}M_{\text{c}} - 2 E_{f} \int \Gamma_{s/v} d^{3}x - \frac{dE_{\text{GW}}}{dt}.
\label{mastereqn}
\end{equation}
The three terms on the right hand side of the above equation correspond to the superradiant growth, production of neutrinos, and emission of gravitational waves respectively.
In the growth phase, the mass of the cloud increases due to the extraction of energy from the black hole through superradiance. However the mass of the cloud is much less than the mass of the black hole, making the angular momentum extraction negligible and thus the spin remains constant in this phase. For scalar bosons the superradiant growth rate can be expressed $\alpha_{g}^{8}\alpha_{*}m_{\phi}/24$ and for the vector case it is $4\alpha_{g}^{6}\alpha_{*}m_{A^{\prime}}$. The mass evolution in this phase can be understood by taking the first term on the RHS of Eq.~\eqref{mastereqn}. This is because, in this phase the fermion production rate is negligible and the superradiant growth rate dominates. Therefore the governing equation takes the form, 
\begin{equation}
\frac{dM_{\text{c}}}{dt} = \Gamma_{\text{SR}}M_{\text{c}}. 
\label{massevolgrow}
\end{equation}
Examining the above equation, one can find the duration for which the superradiant growth takes place. This can be expressed as, 
\begin{equation}
\label{tgrowth}
\tau_{\text{growth}}=\frac{1}{\Gamma_{\text{SR}}}\ln \left(\frac{C_{\phi/ A^{\prime}} \Psi_{c}^{2}}{\alpha_{g}^{3}m_{\phi/ A^{\prime}}}\right),
\end{equation}
where $C_{\phi}=186$ and $C_{A^{\prime}}=\pi$. As the cloud grows during the growth phase, the rate of the energy extraction in the form of fermion production also increases till it balances the energy extraction by the cloud, which marks the beginning of the balanced phase. In the balance phase due to the quenching of superradiance the mass of the cloud remains roughly the same while the black hole spins down up to the threshold mentioned in Eq.~\eqref{spin}. The evolution of the spin is given as,
\begin{equation}
\frac{d a_{*}}{d t}=-\frac{m_{\phi / A^{\prime}}\Gamma_{\text{SR}}M_{\text{c}}}{\alpha_{g}^{2}}.
\label{spinevol}
\end{equation}
Once the spin reduces to a value less than the threshold value, superradiance can no longer sustain and the balance phase ends. The time for which this balance lasts, $\tau_{\text{balanced}}$, can be obtained by solving the above differential equation. 
Then comes the depletion phase, because after superradiance is stopped there will be production of fermions until the peak value of the field is high enough and satisfies the production conditions mentioned in section \ref{subsec:nuprod}. The cloud mass evolution adheres to the equation given as,
\begin{equation}
\frac{dM_{\text{c}}}{dt} = - 2 E_{f} \int \Gamma_{s/v} d^{3}x.
\label{massevoldepl}
\end{equation}
The third term in Eq.~\eqref{mastereqn} corresponds to the loss of energy from the cloud after the depletion phase in the form of gravitational waves. Since the energy loss through this channel is extremely low we do not take this scenario into consideration.
\section{Diffuse Quenched Superradiance Neutrino Background}
\label{sec:dqsnb}
We have introduced the idea of quenched superradiance in the previous section where neutrino production compensates for the colossal growth of the boson cloud. In this article we focus on a diffuse neutrino background created out of these neutrinos which are originated from the quenched superradiance of the black holes. As in the case of the familiar diffuse supernova neutrino background \cite{Beacom:2010kk,Totani:1995rg, Ando:2004hc, Horiuchi:2008jz, Krauss:1983zn}, which is a theoretical prediction of a neutrino signal emanating from all the core-collapse supernova \cite{Colgate:1966ax, Bethe:1985sox, Janka:2016fox, Burrows:2020qrp, Takiwaki:2013cqa} explosions that have happened in the past, here we consider the neutrino flux from the all the primordial black holes instead of focusing only on a definite point source. These black holes were formed in the very early universe i.e., at very high redshifts. This makes the calculation of neutrino energy and flux non-trivial. In the following we discuss the computation of these redshifted observables in detail.
\subsection{Neutrino Background: From Early Universe to Today}
\label{subsec:nubkg}
As seen in section \ref{subsec:nuprod} the energy and flux of the outgoing neutrinos depend on the mass of the cloud-forming boson, the coupling between the neutrino and the bosonic field, the gravitational fine structure constant, and the spin of the black hole. 
Primordial black holes formed in the early universe is central to our study. PBHs satisfying the relevant conditions on spin and conducive to the superradiance phenomena, have a finite time for which the quenched superradiance occurs which is the $\tau_{\text{balanced}}$ explained in section \ref{subsec:timeqsr}. Therefore it can happen that the neutrino production took place very early in the universe. For this reason it is crucial to examine the observed energy and  flux which is redshifted from the energy and flux at the source. The relevant relations are,
\begin{align}
\label{energyz}
E_{\text{obs}} &= \frac{E_{\text{s}}}{1+z},\\
\label{Robs}
R_{\text{obs}} &=\frac{R_{\text{s}}(E_{\text{obs}})}{1+z},
\end{align}
where the cosmological red shift $z$ is defined as the ratio of the scale factor at present $(a_{0})$ to earlier times $(a_{t})$  i.e, $1+z=a_{0}/a_{t}$. Furthermore, in the above expressions, $E_\text{s}$ ($E_{\text{obs}}$) is the energy of the neutrons in the source (observer) frame and $R_\text{s}$ ($R_{\text{obs}}$) is the neutrino flux at the source (observer) frame. As mentioned in the previous section, in the balanced phase of the quenched superradiance, the neutrinos have a fixed energy at the source frame. Therefore, the source frame flux $R_{\text{s}}$ can be expressed as,
\begin{equation}
\label{Rs}
R_{\text{s}}=E_{f}\frac{d \phi}{dE_{\text{s}}}(4\pi d^{2})\delta(E_{\text{s}}-E_{f}),
\end{equation}
here $E_{f}$ is the fixed energy for the neutrino flux  from a single PBH in the source frame and for brevity we denote $R_{\text{tot}}\equiv E_{f}\frac{d \phi}{dE_{\text{s}}}(4\pi d^{2})$. 
Our primary focus is to find the observed diffuse flux. We now have at hand $R_{\mathrm{obs}}$ in units of $\text{eV}^{-1}\text{s}^{-1}$. The diffuse flux in $\text{eV}^{-1}\text{s}^{-1}\text{cm}^{-2}$ can be estimated as,
\begin{equation}
\label{phiobs}
 \frac{d\Phi_{\text{obs}}}{dE}(E_{\text{obs}})=c\int_{z_{\text{lower}}}^{z_{\text{upper}}
} \left| \frac{dt}{dz}\right| \, n_{\text{PBH}}(z)R_{\text{obs}} ~dz.  \end{equation}
Here $z_{\text{upper}}$ is the red shift corresponding to the time when the formation of the PBH and the superradiant growth of the cloud is done. $z_{\text{lower}}$ is the red shift for the time when the formation of the cloud, superradiant growth of the cloud and also the completion of the balanced phase is done. In the above expression $c$ is in $\text{cm}~\text{s}^{-1}$, $\left| dt/dz \right|=1/(1+z)H(z)$ where $H(z)$ is the Hubble parameter at redshift $z$ and
$n_{\text{PBH}}(z)$ is the number density of PBHs at redshift $z$ and is given by, 
\begin{equation}
\label{numden}
n_{\text{PBH}}(z)=\frac{f_{\text{PBH}}\rho_{\text{DM,0}}}{M_{\text{PBH}}}(1+z)^{3},
\end{equation} 
where $f_{\text{PBH}}$ is the PBH abundance, i.e., the fraction of dark matter that is composed of primordial black holes. Although $f_{\text{PBH}}\in[0,1]$, there are various theoretical and (non-)observational bounds on the abundance depending on the PBH mass, and in our article we always consider the maximum allowed value for the different masses of PBH that we consider. The density of dark matter in the present universe given in units of $\text{eV} \text{cm}^{-3}$ is denoted by $\rho_{\text{DM,0}}$ and $M_{\text{PBH}}$ is the mass of a PBH in eV. The number density at the present time is $n_{(\text{PBH,0})}\equiv f_{\text{PBH}}\rho_{\text{DM,0}}/M_{\text{PBH}}$. We use the expression for energy at source, from Eq. \eqref{energyz}, to arrive at the form of the delta function in Eq. \eqref{Rs} as,
\begin{equation}
\label{deltafn}
\delta(E_{\text{s}}-E_{f})=\frac{1}{E_{\text{obs}}}\delta(z-z_{*}),
\end{equation}
here $z_{*}=\frac{E_{f}}{E_{\text{obs}}}-1$.
Simplifying the integral in the expression for flux by plugging in the equations for $R_{\text{obs}}$, black hole number density and delta function from Eqs. \eqref{Robs}, \eqref{numden} and \eqref{deltafn} respectively we obtain,

\begin{equation}\label{finalphi}
\frac{d\Phi_{\text{obs}}}{dE} =
\begin{cases}
    cn_{(\text{PBH,0})}\frac{1+z_{*}}{H(z_{*})}\frac{R_{\text{tot}}}{E_{\text{obs}}} & z_{\text{lower}} \leq z_{*} \leq z_{\text{upper}} \\
    0 & \text{Everywhere else}.
\end{cases}
\end{equation}
Here $z_{\text{upper}}$ is the redshift at the time when the balanced phase of the quenched superradiance was reached and $z_{\text{lower}}$ is the redshift at the time when the depletion phase of the quenched superradiance began.
The Hubble parameter $H_{z_{*}}$ can be calculated as,
\begin{equation}
\label{Hubble}
H_{z_{*}}=H_{0} \sqrt{\Omega_{r}(1+z_{*})^{4}+\Omega_{m}(1+z_{*})^{3}+\Omega_{\Lambda}}.
\end{equation}
$(1+z_{*})$ can be replaced by $\frac{E_{f}}{E_{\text{obs}}}$. $\Omega_{r}$, $\Omega_{m}$ and $\Omega_{\Lambda}$ are the density parameters for radiation, matter and dark energy respectively. $H_{0}$ is the Hubbles's constant.

Hence we obtained the diffuse flux in the range $z_{\text{lower}} \leq z_{*} \leq z_{\text{upper}}$. Substituting for the expression of $z_{*}$ mentioned before, in the range of $z$, we get the diffuse flux energy range to be
\begin{equation}
\label{energyrange}
\frac{E_{f}}{1+z_{\text{upper}}} \leq E_{\text{obs}} \leq \frac{E_{f}}{1+z_{\text{lower}}}.
\end{equation} 
It can be noted here that the energy range given here of the DQS$\nu$B has a large spread for high initial $z$ values. The flux of neutrinos we are considering originated early in the universe at high red shift values. This plays an important role in distinguishing these signals from the diffuse supernova neutrino background (DSNB) which happened much later. The DSNB has a characteristic energy spectrum which has a spread of a few MeVs. 
\subsection{Bounds on Neutrino Background}
\label{subsec:bounds}
In the previous sections we have discussed about the diffuse quenched superradiance neutrino background which are produced through either parametric excitation of oscillating fields or through Schwinger pair production as part of quenching the growth of the bosonic particles formed through superradiance.
Since these neutrinos are emitted at very high red shifts or in early universe the background can be cosmologically constrained. The bounds can also arise from experiments such as the Super-Kamiokande \cite{Super-Kamiokande:2021jaq,Super-Kamiokande:2021the}.

The early universe accommodated relativistic particles such as photons and neutrinos. In the $\Lambda$CDM model, the number of relativistic degrees of freedom in the early universe contributing to the radiation energy density is given by the parameter $N_{\mathrm{eff}}$. The total radiation energy density is given as 
\begin{equation}
\label{rhotot}
\rho_{\mathrm{rad}}=\rho_{\gamma}\left[1+\frac{7}{8}\left(\frac{4}{11}\right)^{\frac{4}{3}}N_{\mathrm{eff}}\right],
\end{equation}
where $\rho_{\gamma}$ is the photon energy density.
A change in $N_{\mathrm{eff}}$ affects the expansion rate of the universe, which in turn would change the time of matter radiation equality. The acoustic peaks of CMB is dependent on the ratio of radiation to matter and any extra radiation would cause a shift in this peak. From this, the Planck (2018) data \cite{Planck:2018vyg} was obtained to be 
$N_{\mathrm{eff}}=2.99 \pm 0.17$ (at $68\%$ CL). On the other hand the standard model predicted value is 3.046 \cite{Mangano:2005cc}. This mismatch in the value of the effective degrees of freedom at the epoch of recombination led to 
$\Delta N_{\mathrm{eff}}= N_{\mathrm{eff}}^{\mathrm{measured}}-N_{\mathrm{eff}}^{\mathrm{SM}}$ that quantifies the excess radiation energy density.
Hence at $95\%$ CL $\Delta N_{\mathrm{eff}}\leq 0.3$. Any generic mechanism, like the one we focus on in this article, which leads to neutrino production in the early universe can explain this excess. This can be expressed as,
\begin{equation}
\label{deltaNgen}
\Delta N_{\mathrm{eff}}=\frac{8}{7}\left(\frac{11}{4}\right)^{\frac{4}{3}} \frac{\rho_{\nu}^{\mathrm{extra,now}}}{\rho_{\gamma}^{\mathrm{now}}}.
\end{equation} 
Here $\rho_{\nu}^{\mathrm{extra,now}}$ is the energy density of the excess radiation today and can be expressed as, 
\begin{equation}
\label{rhonu}
\rho_{\nu}^{\mathrm{extra,now}}=\frac{4 \pi}{c} \int_{0}^{\infty} E\frac{d\Phi_{\mathrm{pre}}}{dE}\,dE.
\end{equation}
It is to be noted here that, since we focus on the excess radiation energy density at the epoch of recombination, any neutrinos produced after that will not contribute to the $\Delta N_{\mathrm{eff}}$ calculation and hence we consider the part of the diffuse flux that was created before the epoch of recombination and is denoted by $\Phi_{\mathrm{pre}}$.
This in turn leads us to the useful expression,
\begin{equation}
\label{deltaN}
\Delta N_{\mathrm{eff}}= 7.083 \times 10^{-9} \int_{0}^{\infty} E\frac{d\Phi_{\mathrm{pre}}}{dE}\,dE.
\end{equation}
Requiring $\Delta N_{\mathrm{eff}}$ to be less than or equal to 0.3 we obtain lower bounds on the scalar and vector couplings. 
Turning to bounds from experiments, we examine constraints from Super-Kamiokande (SuperK). At $90\%$ CL the upper limit on the fluxes for neutrino energies greater than 17.3 MeV is $2.3\, \mathrm{cm}^{-2}\mathrm{s}^{-1}$. Thus by integrating the flux given in \eqref{finalphi} over energy and using the above mentioned constraints we obtain lower bounds on the couplings involved.

\section{Results}
\label{sec:res}
In this section we discuss the implications of this DQS$\nu$B. At first we discuss the DQS$\nu$B with a few benchmark parameters and then we come to the main point of this article, i.e., the bounds on the interaction between the neutrinos and the bosons.

\subsection{Benchmark Parameters}
\begin{figure}[H]
\centering
\includegraphics[scale=0.7]{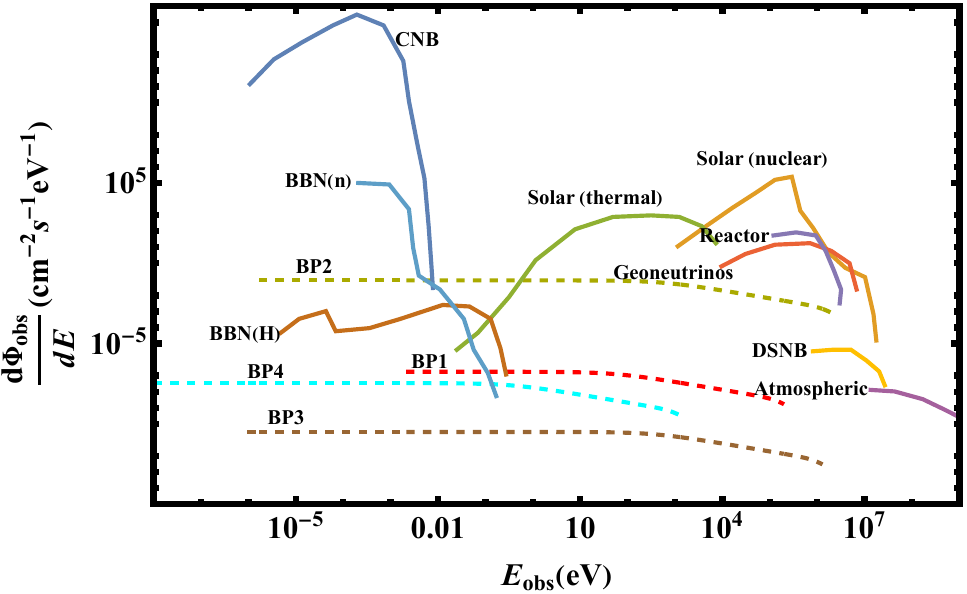}
\caption{The DQS$\nu$B due corresponding to the the benchmark parameters given in Tab.~\ref{bpscavec}. The detected and predicted flux from various other neutrino sources have also been shown for comparison.}
\label{bpplot}
\end{figure}

In order to show the properties of the DQS$\nu$B, we consider a few benchmark cases for both vector and scalar mediated quenched superradiance in top and bottom panel of Tab.~\ref{bpscavec} respectively.
\begin{table}[H]
\centering
\begin{tabular}{|c|c|c|c|c|}
\hline
 & $m_{A^{\prime}}$ (eV) & $g_v$ & $f_{\mathrm{PBH}}$ & $M_{\mathrm{PBH}}$ (g) \\ \hline
BP1 & $10^{-14}$ & $10^{-10}$ & $8.35\times 10^{-7}$ & $7.978\times 10^{35}$ \\ \hline
BP2 & $10^{-6}$ & $3.16\times 10^{-7}$ & $0.062$ & $7.978\times 10^{27}$ \\ \hline
\end{tabular}
\begin{tabular}{|c|c|c|c|c|}
\hline
 & $m_{\phi}$ (eV)~ & $g_s$ & $f_{\mathrm{PBH}}$ & $M_{\mathrm{PBH}}$ \\ \hline
BP3 & $10^{-3~}$ & $~~~~~10^{-4}~~~~$ & $8.3\times 10^{-3}~$ & $7.978\times 10^{24}$ \\ \hline
BP4 & $10^{2~}$ & $~~~~~10^{-2}~~~~$ & $1$ & $7.978\times 10^{19}$ \\ \hline
\end{tabular}
\caption{The benchmark parameter for the vector (top) and scalar (bottom) mediated DQS$\nu$B.}
\label{bpscavec}
\end{table}
The above BPs are motivated by the fact that for these masses of the bosons, these couplings will not lead to the violation of any of the bounds that we discuss in the next section. Using the benchmark parameters in the above table we show the resulting DQS$\nu$B in Fig.~\ref{bpplot}.
The first property that we can see from the vector BPs is that the diffuse neutrino background due to these have a very large energy range. This is due to the fact that the PBHs corresponding to these BPs had reached the balanced phase very early in the universe and hence the energy of the neutrinos got redshifted to much lower values. To illustrate this further, we refer to the DQS$\nu$B flux resulting from the BP4, i.e., the BP corresponding to the lowest PBH mass (and largest boson mass). The energy range for this BP is the widest and neutrinos can in principle be found with energy lower than that of CNB. This is expected because the lightest PBH would have the earliest formation time and heavier bosons would lead to faster growth which results in the activation of this source at a very high redshift. Furthermore, in the figure, only CNB \cite{Weinberg:1962zza,Lesgourgues:2006nd,Mangano:2005cc}, BBN \cite{Wagoner:1966pv,Dolgov:2002wy} and DSNB \cite{Beacom:2010kk,Totani:1995rg,Ando:2004hc,Horiuchi:2008jz,Krauss:1983zn} are the other diffuse background candidates and as we show in this figure, DQS$\nu$B is distinctly different from them as the flux remains constant almost throughout the energy range. This is due to the fact that the other sources of diffuse backgrounds were either active at a very high redshift (CNB, BBN) or they were activated at low redshits (DSNB) whereas due to the prolonged balanced phase of quenched superradiance and the primordial nature of the participating black holes, DQS$\nu$B sources started producing at a very high redshift and were active till low redshift making them one of a kind. Their detection gives us an opportunity to be certain that there exists an exotic mechanism which involves PBHs and BSM degrees of freedom.

\subsection{The bounds on the couplings}
\begin{figure}[H]
\centering
\includegraphics[scale=0.56]{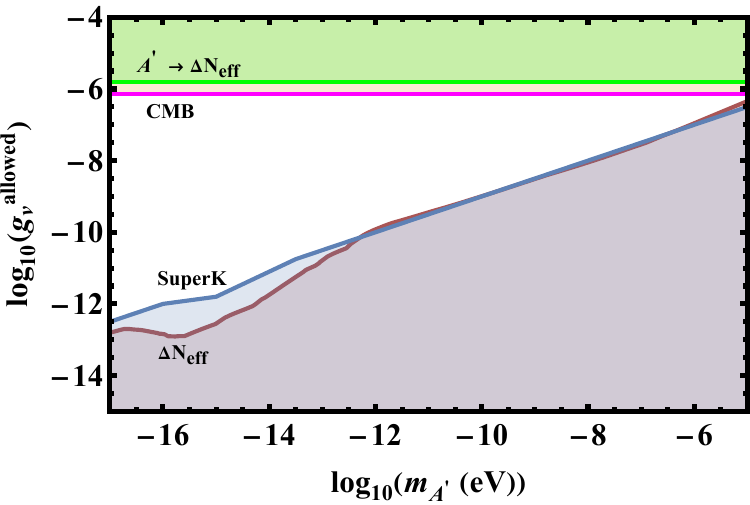}~~~
\includegraphics[scale=0.56]{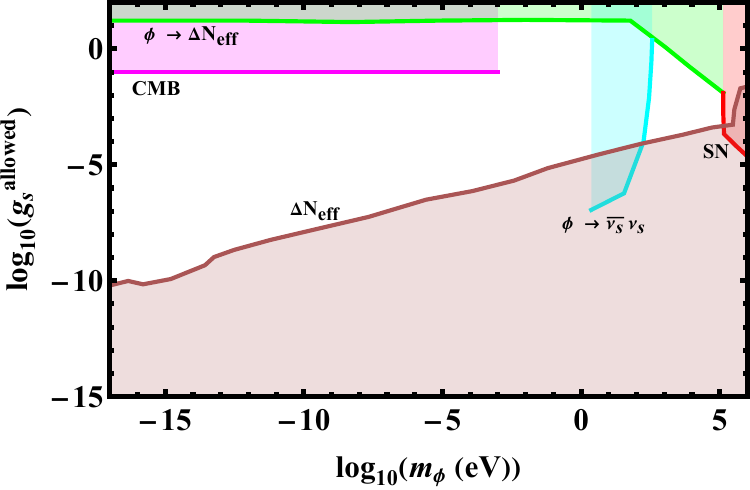}

\caption{(Left) Bounds of $g_v$ from the vector mediated DQS$\nu$B mechanism along with other existing universal bounds. The brown bounds depict the constrains from $\Delta N_{\mathrm{eff}}$ at recombination from neutrinos and the blue bounds depict the constrains from the Super-Kamiokande data.  (Right) Bounds of $g_s$ from the scalar mediated DQS$\nu$B mechanism along with other universal bounds. The brown bounds depict the constrains from $\Delta N_{\mathrm{eff}}$ at recombination from neutrinos. }
\label{bounds}
\end{figure}

As discussed in the previous section, excess radiation energy density and neutrino experiments such as the Super-Kamiokande can be used to put bounds on the DQS$\nu$B, which in turn leads to the bounds on the coupling between the neutrinos and the BSM bosons. The most interesting feature of these bounds are that they are lower bounds, i.e., for a fixed initial spin of PBHs one can limit these couplings from below depending on the mass of the boson. We show these bounds in Fig.~\ref{bounds}.
In the left panel of the above plot we also show the universal upper bounds on the active neutrino-vector boson coupling from the CMB data \cite{Sandner:2023ptm} in magenta and from the contribution of the vector in $\Delta N_{\mathrm{eff}}$ at BBN in green \cite{Huang:2017egl}. In the right panel of the plot, we show the universal upper bound on the sterile neutrino-scalar boson coupling from the CMB data in magenta \cite{Forastieri:2019cuf}, from the contribution of the scalar in $\Delta N_{\mathrm{eff}}$ at BBN in green \cite{Huang:2017egl}, from the scalar decay into neutrino pair \cite{Escudero:2019gvw} in cyan and from supernova in red \cite{Brune:2018sab,Heurtier:2016otg}. 
As mentioned before, for the scalar case, our bounds are applicable for the interaction strength between the sterile neutrino and the BSM scalar. Therefore, in showing the universal upper bounds in the figure we have converted the bounds on the interaction strength between active neutrino-BSM scalar ($g_{s,\mathrm{active}}^{\mathrm{bound}}$) using the relation $g_s^{\mathrm{bound}} \sim g_{s,\mathrm{active}}^{\mathrm{bound}}/|U_{s\alpha}|^2$, where $U_{s\alpha}$ is the mixing between sterile and active neutrino which we consider to be $10^{-3}$. Furthermore, in both the cases, we have shown the bounds for the boson mass range where the highest boson mass is determined by the overlap of the upper and the lower bounds of the coupling constrains and the lowest value of boson mass corresponds to the supermassive black hole with mass $\mathcal{O}(10^{6})M_{\odot}$.

In both the cases of scalar and vector mediated quenched superradiance, we consider the initial spin to be $a_* = 0.9$ and the gravitational fine structure constant $\alpha_g = 0.3$, which fixes the mass of the PBH corresponding to the mass of the boson. Under these considerations, we show the lower bounds on the couplings. It is to be noted that these bounds appear in the form of lower bounds because as the coupling reduces, the field value for the balance condition increases which in turn increases the flux of neutrinos. For the vector case we find that both the bounds from Super-Kamiokande and the bounds from $\Delta N_{\text{eff}}$ are relevant whereas for the scalar case only the latter is applicable.

It is worth mentioning here that changing the initial spin of the PBH and the gravitational fine structure constant of the gravitational atom would alter the bounds which can be easily obtained from the prescription given in the previous section. On the other hand, we have only considered monochromatic mass and spin spectra of the PBH for simplicity whereas considering non-monochromaticity in both these cases would lead to non-trivial changes in the bounds which we leave for future work. Furthermore, one may consider complete models with more complicated interaction terms between the neutrinos and the bosons that may lead to model specific bounds.

\section{Conclusion}
\label{sec:concl}
In this article we have considered a novel diffuse background of neutrinos which originate from the quenched superradiance of primordial black holes and we term this background as diffuse quenched superradiance neutrino background. At first we have briefly discussed the quenched superradiance mechanism, the neutrinos flux and energy originating from this mechanism and the relevant timelines of this mechanism. Next we discussed the evolution of spinning PBHs under the influence of quenched superradiance leading to the DQS$\nu$B. 

We show that for specific initial spin, boson mass and coupling between neutrinos and bosons, a novel neutrino diffuse background can originate which has properties that distinguishes it from the already theorised neutrino backgrounds such as the CNB, BBN, and DSNB. We find that DQS$\nu$B usually has larger energy range and the flux stays (nearly) constant over the entire energy range. This make the neutrinos in principle detectable through future experiments with increased sensitivity at various different energy ranges. Finally we moved to the main aim of this article, i.e., in this work for the first time, we put lower bounds on the neutrino and BSM boson coupling using the DQS$\nu$B. For the vector mediated DQS$\nu$B we find that both $\Delta N_{\mathrm{eff}}$ and Super-Kamiokande can put lower bounds on the BSM vector-active neutrino coupling whereas for the scalar mediated DQS$\nu$B we manage to put lower bounds on the BSM scalar-sterile neutrino coupling from $\Delta N_{\mathrm{eff}}$ bound.

It is worth mentioning that, in developing the prescription to the novel neutrino background, we have made some simplifying assumptions. For the scalar case, since active neutrinos will not lead to an efficient quenched superradiance, we consider sterile neutrino-BSM scalar interaction and then consider the neutrino oscillation during propagation to obtain the active neutrino flux. However, we consider an active-sterile neutrino mixing of $\mathcal{O}(10^{-3})$, and different values of this mixing will alter the bounds. On the other hand, for both scalar and vector cases we assume monochromatic spectra of mass and spin. However non-monochromaticity in spin and mass would lead to non-trivial changes in the bound. Furthermore, we have obtained our bounds for fixed values of initial PBH spin and gravitational fine structure constant and changing these would lead to some changes in the bounds. Finally, one might consider complete models to obtain model dependent bounds.

In conclusion, in this article, we for the first time give prescription to obtain a novel neutrino background due to quenched superradiance of primordial black holes with unique features. Detection of this kind of a background will open new doors in the process of BSM investigation.

\acknowledgments{UKD acknowledges support from the Anusandhan National Research Foundation (ANRF), Government of India under Grant Reference No.~CRG/2023/003769. IKB thanks Mr. Rajib Maity and Mrs. Priya Maity for their incredible hospitality. AJ thanks the DST-INSPIRE for support through the INSPIRE fellowship.}


\bibliographystyle{JHEP}
\bibliography{DQSNB}

\end{document}